\documentclass{jfm}
\usepackage{amsmath,bm,amssymb,amsfonts}
\usepackage{natbib}
\usepackage{graphicx}
\usepackage{color}
\definecolor{RED}{rgb}{0.6,0.,0.}
\definecolor{BLUE}{rgb}{0.,0.,0.6}
\definecolor{GREEN}{rgb}{0.,0.6,0.}

\newlength{\picwidth}
\newlength{\plotwidth}
\begin{document}

\title{Reactive Rayleigh-Taylor Turbulence}

\author{M. Chertkov$^1$, V. Lebedev$^{1,2}$, and N. Vladimirova$^{1,3,4}$}

\affiliation{$^1$ Theoretical Division \& CNLS, LANL, Los Alamos, NM 87545, USA \\
 $^2$ Landau Institute for Theoretical Physics RAS \\
 $^3$ ASC Flash Center, University of Chicago,
 5640 S Ellis Ave, Chicago, IL 60637, USA \\
 $^4$ Department of Mathematics and Statistics, UNM,
   Albuquerque, NM 87131, USA 
}

\maketitle

\begin{abstract}
The Rayleigh-Taylor (RT) instability develops and leads to turbulence
when a heavy fluid falls under the action of gravity through a light
one. We consider this phenomenon accompanied by a reactive
transformation between the fluids, and study with Direct Numerical
Simulations (DNS) how the reaction (flame) affects the turbulent
mixing in the Boussinesq approximation. We discuss ``slow" reactions
where the characteristic reaction time exceeds the temporal scale of
the RT instability, $\tau\gg t_\mathrm{inst}$. In the early turbulent
stage, $t_\mathrm{inst} \lesssim t\lesssim\tau$, effects of the flame
are distributed over a maturing mixing zone, whose development is
weakly influenced by the reaction. At $t\gtrsim\tau$, the fully mixed
zone transforms into a conglomerate of pure-fluid patches of sizes
proportional to the mixing zone width. In this ``stirred flame''
regime, temperature fluctuations are consumed by reactions in the
regions separating the pure-fluid patches. This DNS-based qualitative
description is followed by a phenomenology suggesting that thin
turbulent flame is of a single-fractal character, and thus
distribution of the temperature field is strongly intermittent.
\end{abstract}


\section{Introduction}

The Rayleigh-Taylor (RT) instability \cite[]{1883Ray,50Tay,61Cha}
occurs in many natural and man-made flows. Under constant
acceleration, produced e.g. by gravity, the instability develops into
the RT turbulence \cite[]{62DHH,84Sha}, characterized by a mixing zone
which continuously grows.  Many recent experimental
\cite[]{99DLY,02WA}, numerical
\cite[]{01YTDR,99DLY,01CD,04RC,06CC,07VC} and theoretical efforts
\cite[]{03Che,05CKL} --- later ones in the spirit of the turbulence
phenomenology \cite[]{41Kol,49Obu,51Cor,95Fri} --- have been devoted
to analysis of turbulence inside the mixing zone. The mixing fluids
can also be involved in chemical or nuclear reactions, e.g. flame. A
flame front propagating against the acceleration is modified by the RT
instability, which leads to a reactive turbulence in the mixing
zone. Studies of the externally stirred turbulent flames belong to the
general field of turbulent combustion
\cite[]{40Dam,85Wil,Peters,05PV,Borghi,01Ker} and have many important
applications.  The buoyancy driven, reactive turbulence is believed to
be the dominant mechanism for thermonuclear burning in type-Ia
supernovae \cite[]{95Kho,03GKOCR,05ZWRDB}.  The interplay of buoyancy,
reactive transformation and turbulence also plays a crucial role in
studies of fusion \cite[]{77FCT} and large-scale combustion, such as
furnaces and fires \cite[]{96CetKas,01Tie}. Turbulence, generated by
buoyancy, facilitates mixing, and thus counteracts the dynamical
separation of fluids (phases) due to reaction. To explain this
competition between {\bf separation and mixing} is the main challenge
emerging in the description of reactive flows \cite[]{01Ker}, which
also applies in the broader context of multi-phase fluid mechanics,
e.g. the RT turbulence of an immiscible mixture controlled by surface
tension \cite[]{05CKL}.

To clarify the nature of the competition between separation and
mixing, let us consider a typical reactive RT setting.  Initially, the
heavy, cold reactant is placed on top of the light, hot product. As in
the non-reacting RT instability, the mixing zone develops.  Since
locally the reaction occurs only in the mixed fluid, turbulence inside
the mixing zone enhances the cumulative reaction rate. The effect of
reaction, however, is the opposite -- the reaction consumes mixed
material, and potentially limits the growth of the mixing zone. While
diffusion and turbulence mix the reactant with the product, the
reaction converts the mixed fluid into the product, thus causing an
upward shift of the mixing zone as a whole.

Existing studies of reactive RT setting have mainly focused on the
initial stages of instability \cite[]{06LSW}, or on the instability
restricted by container walls in experiments \cite[]{04BM} and by
domain sizes in simulations \cite[]{95Kho,05ZWRDB,05VR}. The transverse
(to direction of gravity) confinement results in a constant (on
average) cumulative reaction speed, presumably dependent on the domain
size \cite[]{95Kho}. In this system fluctuations, e.g. in velocity and
density fields, do not grow with time. It is still not clear whether
reaction can stabilize the unconfined RT instability. One suspects
that the answer might depend on the relation between the timescale of
the RT instability and the reaction timescale.

In this paper, we study RT turbulence unconfined by walls in the
presence of reactions slower than RT instability.  Armed with DNS and
building a unifying phenomenological description, we analyze the
interplay of turbulent mixing and reaction-mediated
separation. Specifically, we focus on predicting macroscopic and
microscopic features of the mixing zone, such as the upward transport
of the mixing zone as a whole and statistics of density fluctuations
within the mixing zone.

We examine the interacting effects of buoyancy and reaction in the
simplest, however physically relevant setting, the Boussinesq
approximation, where the variations in the fluid density are small and
are related linearly to the temperature contrast. We assume that the
pressure increase due to reaction is negligible, and that the Mach
number is small and, consequently, the fluid is incompressible. Thus,
our system can be modeled by the incompressible Navier-Stokes equation
coupled to the advection-reaction-diffusion equation governing the
temperature evolution. The set of equations is written as
\begin{eqnarray}
    && \label{NSeq}
    \partial_t {\bm v}+({\bm v}\nabla){\bm v} =
       -\nabla p+\nu\nabla^2{\bm v} - 2{\cal A} \bm g \theta,
   \qquad \nabla{\bm v}=0,
   \\   && \label{Teq}
    \partial_t \theta+({\bm v}\nabla)\theta =
       {\tau}^{-1} R(\theta)+\kappa\nabla^2 \theta,
\end{eqnarray}
where $\bm v$ is the flow velocity, $p$ is the pressure, and $\theta$
represents temperature variations and/or the chemical composition of
the fluid. It is convenient to choose $\theta$ equal to zero in the
cold (heavy) phase and to unity in the hot (light) phase. Often,
$\theta$ is referred to as the reaction progress variable. The
reaction and production of heat is represented by the first term on
the right hand side of Eq.~(\ref{Teq}); we refer to $R(\theta)$ as the
reaction rate function, and to $\tau$ as the characteristic reaction
time, or reaction time scale. We consider a single step reaction
\cite[see][]{05PV}, where the reaction rate is zero in the pure phases,
i.e.  $R(0)=R(1)=0$ and $R=O(1)$ for $0<\theta<1$. The buoyancy force,
i.e. the last term on the right hand side of Eq.~(\ref{NSeq}), is
directed along the gravitational acceleration $\bm g$. The factor
${\cal A} = (\rho_1 - \rho_2)/(\rho_1 + \rho_2)\ll 1$ is the Atwood
number, where $\rho_1$ and $\rho_2$ are densities of the reactant and
the product, respectively. We assume that the kinematic viscosity
$\nu$ in Eq.~(\ref{NSeq}) and the diffusion coefficient $\kappa$ in
Eq.~(\ref{Teq}) are comparable, and thus the Prandtl number is of the
order of unity.

We compare our simulations with the benchmark non-reacting case,
corresponding to $\tau=+\infty$ in Eq.~(\ref{Teq}).  Phenomenological
theory introduced by \citeauthor{05CKL} (\citeyear{03Che,05CKL})
and hereafter called the
``phenomenology'' suggests that in the non-reacting RT turbulence,
velocity and density fluctuations are driven by buoyancy at the scales
$L_v$ and $L_\theta$, respectively, both comparable to the mixing zone
width, $H$. These fluctuations generate cascades (of energy and
temperature) towards small scales. The cascades terminate at the
viscous scale, $\eta$, and (thermal) diffusive scale, $r_d$,
respectively \cite[]{03Che,05CKL}. The scales $\eta$ and $r_d$ are of
the same order provided the Prandtl number is of order
unity. Previously conducted simulations of the non-reactive RT
turbulence in the Boussinesq regime \cite[]{07VC} show that while the
mixing zone width grows in time, $\eta$ and $r_d$ decrease with
time. It was also found 
that the correlation radii of
both the velocity and temperature fluctuations, $L_v$ and $L_{\theta}$
grow in time as $0.05$--$0.1 H$. One special focus 
was on resolving the internal structure of the
mixing zone more systematically than in the previous studies. It was
shown that spatial correlations do not depend much on the vertical
position within the mixing zone. In particular, this means that the
relevant values of the major scales characterizing a snapshot of the
RT turbulence depend only weakly on the height of the horizontal slice
within the mixing zone.

\section{Preliminary Considerations}
\label{Prelim}

In this paper we discuss the regime of relatively slow reaction: the
time scale of the non-reacting RT instability, $t_\mathrm{inst}$, and
the characteristic reaction time, $\tau$, are assumed to be well
separated, $\tau \gg t_\mathrm{inst}$.

At the early stage, $t<\tau$, development of the mixing zone and the
range of turbulent scales inside the mixing zone are weakly influenced
by the reaction. Nevertheless a cumulative effect of the reaction can
be seen in the overall shift of the mixing zone, as a whole, from the
product side to the reactant side. To estimate the shift of the mixing
zone, $z_f$, let us integrate Eq.~(\ref{Teq}) along $z$. For the first
term on the left hand side one arrives at $\partial_t z_f$, while the
only other nonzero contribution is associated with the reaction term
$R/\tau$ on the right hand side.  Since in this regime the heat
production is determined mainly by the mixing on the scale of the
whole mixing layer, the reactive contribution is estimated simply as
$H/\tau$. Thus, $\partial_t z_f\sim H /\tau$. Even though the overall
shift of the mixing zone at $t<\tau$ is smaller than $H$, $z_f$ grows
faster than $H$, so that the two scales become comparable at
$t\sim\tau$. From this time onward the reaction effects become more
prominent.

The reaction becomes more effective when the turbulent turnover time
at $H$, estimated as $t$, exceeds the characteristic reaction time
$\tau$. In this regime the reactant entrained into the mixing zone is
burned out completely, and the overall shift of the mixing zone,
$z_f$, should be of order $H$. This transition in the mixing zone
evolution is also accompanied by a qualitative modification of the
temperature fluctuations in the interior of the mixing zone. In
contrast, the velocity distribution at all the scales is expected to
show only weak sensitivity to the reaction even at $t>\tau$.

The two stages observed for $t<\tau$ and $t>\tau$ will be called the
{\it mixed stage} and the {\it segregated stage}, respectively. In the
field of turbulent combustion \cite[see][]{85Wil,Peters}, the mixed regime
is called ``well-stirred reactor'' \cite[]{05PV}, ``thickened flame"
\cite[]{Borghi} or ``broken reaction zone" \cite[]{Peters} while the
segregated regime is the regime of ``distributed reaction zones''
\cite[]{05PV}, ``wrinkled-thickened flame" \cite[]{Borghi} or ``thin
reaction zone" \cite[]{Peters}. Transition between the regimes
occurring at $t\approx\tau$ can be interpreted as crossing the
$\mathrm{Da}=1$ line on the turbulent combustion diagram
\cite[see][]{05PV}. Indeed, the Damk\"oler number \cite[]{40Dam},
$\mathrm{Da}$, is defined as the ratio of the integral time scale to
the reaction time scale, $\mathrm{Da}=t/\tau$, which grows linearly in
time for the RT turbulence. The other popular dimensionless number in
turbulent combustion, the turbulent Karlovitz number, $\mathrm{Ka}$,
describes transition between the regime of ``distributed reaction
zones''/``wrinkled-thickened flame''/``thin reaction zone'' and the
``flamelet'' regime.  The Karlovitz number is defined as the ratio of
the chemical time scale to the Kolmogorov time scale. For the slowly
reacting RT turbulence studied here, the ``flamelet'' regime does not
apply, i.e. $\mathrm{Ka}$ is always much larger than unity and it
increases with time. The laminar flame regime is also not observed in
the systems considered here; consequently the laminar flame thickness
$h_\mathrm{lam}$ is an irrelevant scale.
  
\begin{figure}
  \begin{center} 
    \includegraphics[scale=0.4]{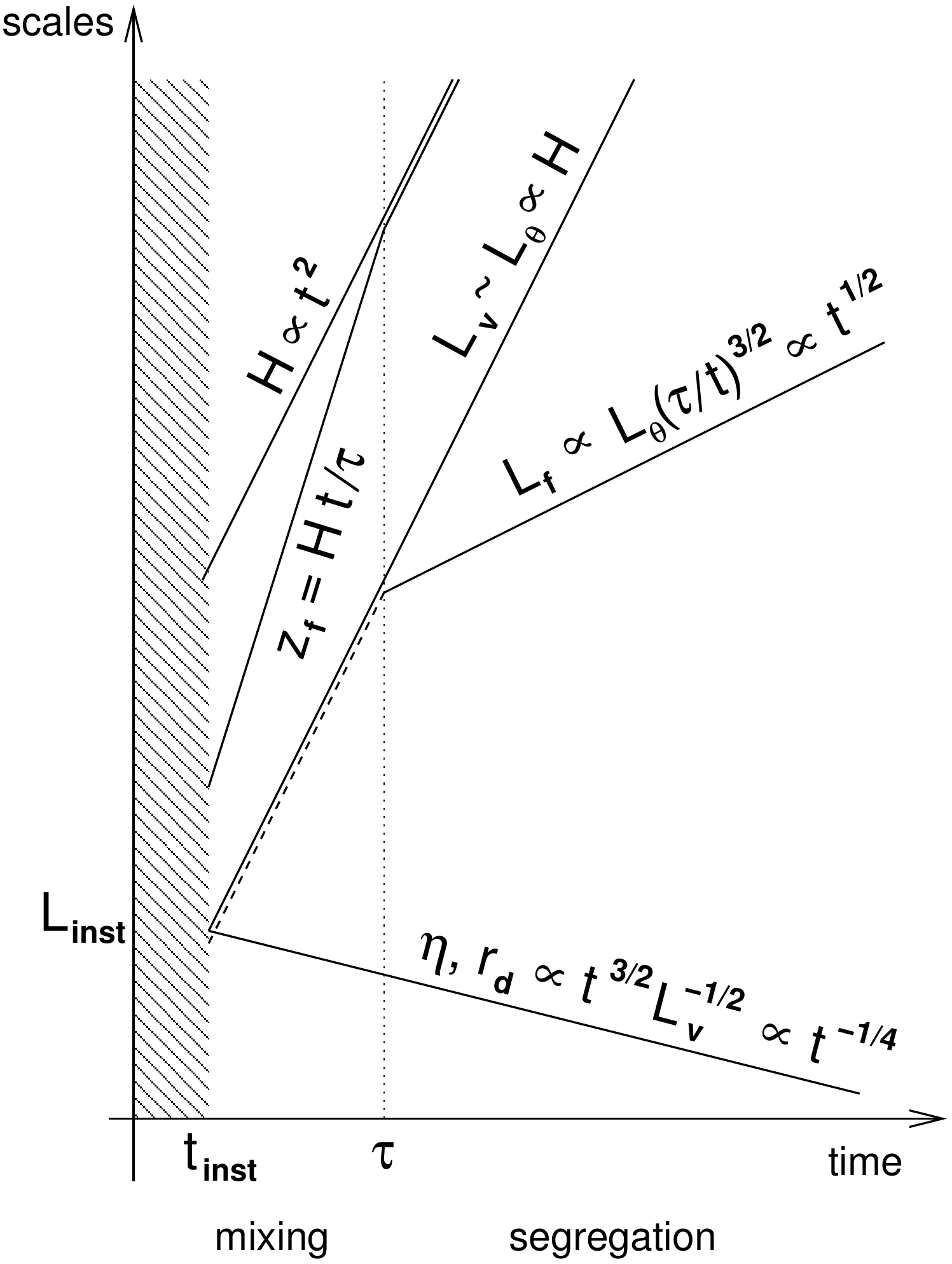}
  \end{center} 
  \caption{ 
  Schematic representation of the system
  evolution in log-log coordinates: the width of the RT mixing layer
  $H$, the large scale fluctuations of velocity and temperature $L_v$
  and $L_\theta$, the shift of the mixing layer $z_f$ due to reaction,
  the turbulent flame thickness $L_f$, and the dissipation scales
  $\eta$ and $r_d$.  The stages of the early RT instability
  development and transition to turbulence are shown shaded. $\tau$ is
  the characteristic reaction time, separating well-mixed and
  segregated regimes. In the mixed regime, the turbulent flame
  thickness corresponds to the typical width of interfaces, separating
  pure phase domains of size~$L_\theta$. The temporal scalings are
  phenomenological propositions discussed in
  Sections~\ref{Prelim},~\ref{Pheno}.  }
  \label{fig:weak} 
\end{figure}

At $t>\tau$ the main effect of the reaction on the mixing zone is in
creation of domains of pure phases ($\theta=0$ and $\theta=1$). In
these domains the fluctuations in the reaction term of
Eq.~(\ref{NSeq}) dominate fluctuations in the respective advection
term, leading to complete suppression (burning out) of temperature
fluctuations. The domains are separated by relatively thin (compared
to domain size) interfaces where burning occurs. The interface width,
also referred to as the turbulent flame thickness, $L_f$, (that is the
scale where the turnover time of velocity fluctuations is comparable
to $\tau$) becomes much smaller than $L_v$ at $t\gg\tau$ due to the
dominance of reaction.

Formation of the conglomerate of pure reactant and product domains
should result in a qualitative transformation of the single-point
statistics of the temperature field inside the mixing zone. A single
peak distribution centered around $\theta=1/2$ transforms into a
distribution with two peaks, related to the emergence of pure phase
domains corresponding to $\theta=0$ and $\theta=1$. The qualitative
differences in the projected stages are summarized in
Fig.~\ref{fig:weak} where the temporal behavior of different
characteristics is shown schematically.

\section{Numerical Results}

\begin{figure}
    \begin{center}
    \includegraphics[width=96mm]{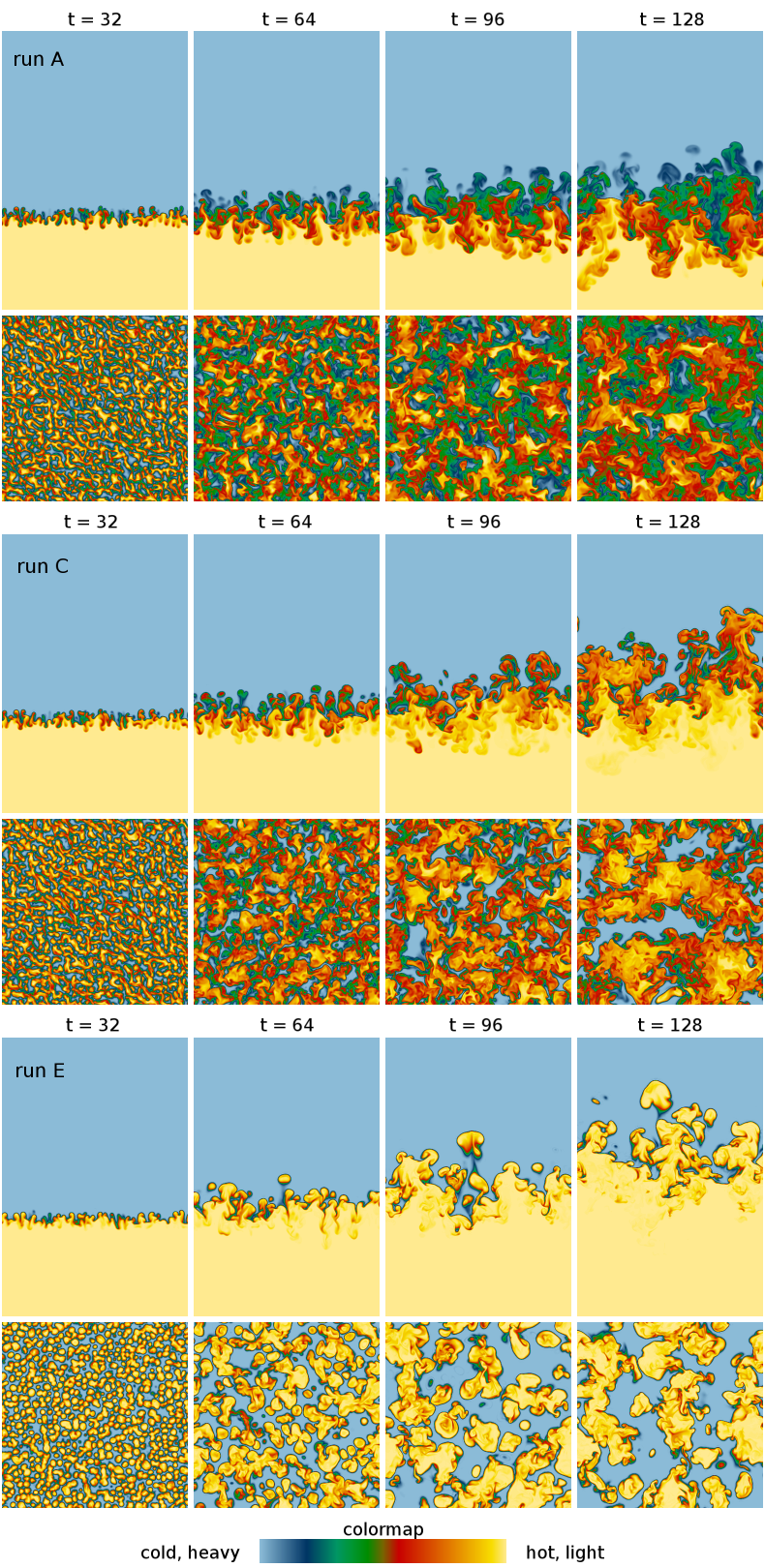}
    \end{center}
    \caption{Vertical and horizontal ($z=z_f$, center of the
             mixing zone) slices of temperature
             for three different reaction strength.}
    \label{T-slices}
\end{figure}

The computational technique used in this work is similar to that described in 
\cite{07VC}. In our simulations, we use the popular Kolmogorov-Petrovskii-Piskunov
(KPP) model \cite[]{Fisher,KPP}, where $R=4\theta(1-\theta)$ with $0<\theta<1$. The KPP
model is not the only possible choice of the function $R(\theta)$. However, compared to
other models, the KPP reaction provides a relatively thick laminar front which can be
simulated at coarser resolution and makes computations more affordable. As 
before, we restrict ourself to the case of $\mathrm{Pr}=1$, and chose the units
where $\nu=\kappa=1$ and $2{\cal A}g=1$. The laminar flame parameters for five simulation
sets discussed below are shown in Table~\ref{tbl}.

\begin{table}

\begin{center}
 \begin{tabular}{crrr}

    \makebox[8mm]{run} &
    \hspace{0mm} $\tau$ \hspace{16mm}&
    \hspace{6mm} $h_\mathrm{lam}$ \hspace{5mm}&
    \hspace{4mm} $s_\mathrm{lam}$ \hspace{4mm}\\
                                             \\
A &  1600.00 &   160.00 & 0.10 \\

B &   177.78 &    53.33 & 0.30 \\

C &    64.00 &    32.00 & 0.50 \\

D &    40.32 &    25.40 & 0.63 \\

E &    16.00 &    16.00 & 1.00    \\

\end{tabular}
\end{center}

\caption{Laminar flame parameter for simulations discussed in the text: reaction time
$\tau$, laminar flame thickness $h_\mathrm{lam}=4\sqrt{\kappa\tau}$, and laminar flame
speed, $s_\mathrm{lam}=4\sqrt{\kappa/\tau}$, for KPP reaction.}

\label{tbl}
\end{table}

We solved the equations (\ref{NSeq},\ref{Teq}) using the spectral
element code developed by 
\cite{code}. We use elements of size $30^3$ with 12
collocation points in each direction. The size of our computational
domain is $960\times 960\times 1440$ physical units, or $384\times
384\times 576$ collocation points. We stop our simulation at time
$t=128$, when the width of the mixing layer approaches the size of the
computational domain. The boundary conditions are periodic in the
horizontal ($x,y$) directions and no-slip in vertical ($z$)
direction. The initial conditions, taken at $t=0$, include a quiescent
velocity and a slightly perturbed interface between the domains,
determined by $\theta=\theta_0(z+\delta)$. The function
$\theta_0(z)=0.5 [1+\tanh(0.4z)]$ describes the density profile across
the interface and $\delta(x,y)$ is a perturbation having the spectrum
with modes $18 \le n \le 48$ with spectral index~0.  Here the spectral
index refers to the exponent of the wave-number 
\cite[see][]{05RDA}, and it describes the shapes of the perturbation
spectra.

The least processed results of numerical simulations, snapshots of the
temperature field $\theta$, are shown in Fig.~\ref{T-slices}; they are
quite informative for our intuition.  In case~A ($\tau=1600$) the
structure of the mixing layer is practically indistinguishable from
the non-reacting case \cite[see][]{07VC}, at least up to
$t=128$. In case~C ($\tau=64$) the reaction starts to influence the
temperature distribution somewhere between $t=64$ and $t=96$, where we
observe emergence of regions of pure phases. In case~E ($\tau=16$)
reactants and products are well separated by relatively thin
interfaces at $t=32$. Thus, a transition from the mixed stage to the
segregated stage occurs at $t\approx\tau$, in accordance with the
preliminary discussion of Section~\ref{Prelim}.

\begin{figure}
 \begin{center}
 \includegraphics[width=120mm]{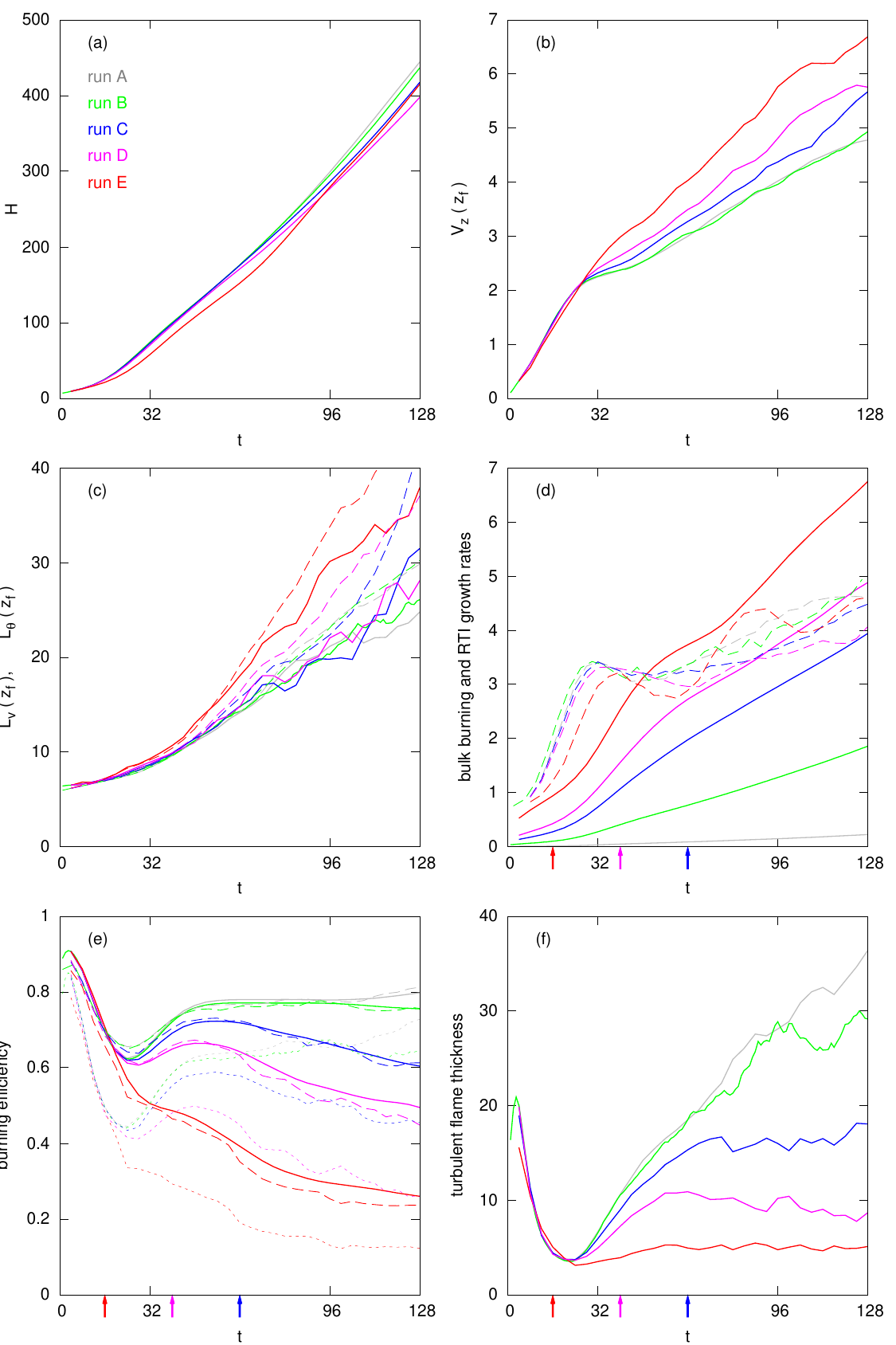}
 \end{center}
 \caption{ 
   (a) half-width of mixing layer $H$; (b) RMS of vertical
   velocity at $z_f$; (c) pumping scale $L_v$ (dashed lines)
   and scale $L_{\theta}$ (solid lines) measured at $z_f$;
   (d) bulk burning rate
   $\dot{z}_f$ (solid lines) compared to growth rate of RT instability
   $\dot{H}$ (dashed dots);  (e) burning efficiency measured as
   ratio $\dot{z}_f \tau /H$ (solid lines), the average reaction rate
   $\bar{R}(z_f)$ in the middle slice (dashed lines), and well-mixed fraction of the
   slice area, $\Delta_{\rm mix}$ (dotted lines);  (f)
   turbulent flame thickness in the middle slice obtained using
   ``expanding circle'' technique. Arrows in (d)-(f) show the reaction time.} 
\label{fig_velocity}
\label{fig_fspeed}
\end{figure}

In Fig.~\ref{T-slices} we observe that the width of the mixing layer,
defined as the vertical distance between bubbles and spikes, is
approximately the same in all cases.  Measurements of the half-width
of the mixing layer, defined as $H=\int R(\bar{\theta})\, dz$ (where
the overbar indicates averaging in a horizontal plane), confirm that
$H(t)$ is weakly dependent on the reaction rate (see
Fig.~\ref{fig_velocity}a). In Fig.~\ref{fig_velocity}b we plot the
root mean square (RMS) of the vertical velocity and see that in all
five cases the RMS velocity is practically the same. Moreover, all
five cases exhibit similar internal structure of the velocity field
inside the mixing layer, as we see from the comparison of the
temperature and velocity correlation lengths, $L_{\theta}$ and $L_v$,
measured in the central slice of the mixing zone
(Fig.~\ref{fig_velocity}c). Following 
our earlier approach \cite[]{07VC}
we define the velocity correlation length, $L_v$, as the half-width of
the the normalized two-point pair correlation function of velocity,
$f(0)/f(L_v)=2$, with $f(r)=\langle v_z(\bm r_1)v_z(\bm r_1+\bm
r)\rangle/\langle v_z^2 \rangle$. The temperature correlation function
is defined in a similar way, but based on the temperature
fluctuations, $\theta - \bar{\theta}$. Both temperature and velocity
correlation lengths differ by 20-30\% compared to each other for
$\tau$ ranging from $16$ to $1600$. Similarly to what was reported in
\cite{07VC} for the non-reacting case, $L_{\theta}/H \approx L_v/H
\approx 0.1$. The viscous length $\eta = (\nu^2/ \langle 5 |\nabla
v|^2 \rangle )^{1/4}$ decreases slowly with time in all cases, with
less than $10\%$ variation between $\tau=16$ and $\tau=1600$. The main
conclusion here is that the velocity field is essentially insensitive
to the reaction in the slow reaction regime.

%

Next we look at the shift of the mixing layer $z_f= \int
\bar{\theta}\, dz$, introduced in Section \ref{Prelim}. We compute the
cumulative reaction, or bulk burning rate, $\dot{z}_f = \int
\overline{R(\theta)} dz$, and compare it to the growth rate of the
mixing layer, $\dot H$. Both quantities are shown as functions of time
in Fig.~\ref{fig_fspeed}d. We see that the bulk burning rate is
greater for higher reaction rates. The time when $\dot H$ and $\dot
z_f$ become comparable is approximately equal to $\tau$.

It is not surprising that the bulk burning rate increases with
decrease in $\tau$.  However, in addition to this obvious effect, we
need to take into account the reaction efficiency, which depends on
how well the fluids are mixed within the mixing layer. In
Fig.~\ref{fig_fspeed}e we show reaction efficiency of the whole mixing
layer, $\dot{z}_f \tau /H$, and compare it to the reaction efficiency
in the middle slice, $\bar{R}(z_f)$.  Good agreement observed between
the two characteristics suggests that the fraction of well-mixed fluid
within the mixing zone does not vary significantly with the vertical
position. In fact, we had also made this observation earlier in our
study of the non-reacting case \cite[]{07VC}, where the mixing
efficiency stayed at the constant level of $\approx 0.8$ across the
whole layer. In addition, Fig.~\ref{fig_fspeed}e shows the fraction of
well-mixed fluid, $\Delta_\mathrm{mix}$, defined as a relative area of
the slice $z=z_f$ occupied by fluid with $0.25 <\theta <0.75$. We see
that $\Delta_\mathrm{mix}$ behaves similar to $\bar{R}(z_f)$; some
differences reflect an arbitrary choice of the selected limits.  The
area fraction $\Delta_\mathrm{mix}$, as well as the average reaction
rate in the middle slice and the reaction efficiency characterize the
transition from the mixed stage to the separated stage. In the mixed
stage $\Delta_\mathrm{mix}$ is of order unity, and in the segregated
stage it is a decreasing function of time. Numerical results shown in
Fig.~(\ref{fig_fspeed}e) confirm a decrease of $\Delta_\mathrm{mix}$
observed for $t>\tau$.

\begin{figure}
\begin{center}
\includegraphics[width=124mm]{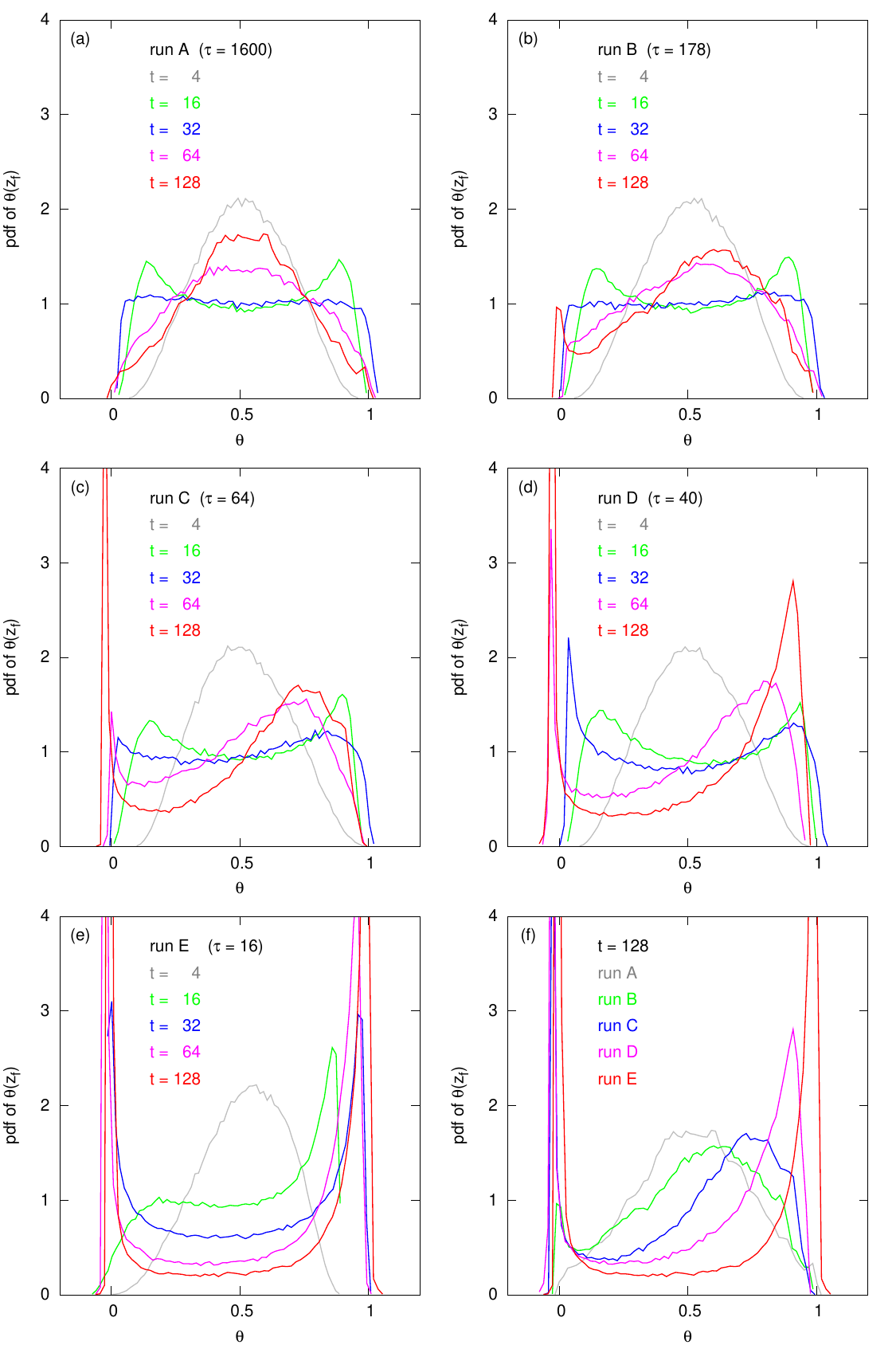}
\end{center}
\caption{PDF of temperature in the middle slice. Panels (a)-(e)
   show time evolution of PDFs.  Panel~(f) compares PDFs at $t=128$
   for different reaction times, $\tau$.
   }
\label{fig_pdfpeaks}
\end{figure}

Since the reaction at the central, $z=z_f$, slice describes the bulk
burning rate so well, we pay particular attention to the temperature
distribution within the slice, shown in Fig.~\ref{fig_pdfpeaks}. In case
A the probability distribution function (PDF) of temperature evolves
in time almost as if there were no reaction at all. It starts from a
single peak distribution determined by the initial conditions
(slightly perturbed diffused interface). Then, during the linear stage
of the RT instability, $t\lesssim 16$, the PDF transforms into the
two-peak shape and becomes single-peaked again at $t\approx 48$. The
transformation from one to two peaks corresponds to the transition to
the non-linear regime of the RT instability, associated with secondary
Kelvin-Helmholtz type shear instability and formation of RT mushrooms,
while transition from two peaks to one corresponds to the destruction
of RT mushrooms and formation of the turbulent mixed zone. When
reaction is stronger, as in cases B and C, we observe yet another
transformation in the PDF, namely the appearance of a narrow peak
around $\theta=0$ and decrease in the middle of the $\theta$ range at
$t\approx\tau$. The transition to the two-peaked shape at
$t\approx\tau$ is of interest to us since it indicates transition to
the segregated regime. (In cases D and E, this 1-2-1-2 peak pattern is
not noticeable because $\tau$ is of the order of, or shorter than, the
transition time from linear to non-linear RTI.) Quantitatively the
process of the transition to the segregated regime can be described by
integrating the PDF function over some interval in the vicinity of
$\theta=1/2$. Selecting the interval $0.25 < \theta<0.75$, we obtain
$\Delta_\mathrm{mix}$, described earlier.

The well-mixed fraction, $\Delta_\mathrm{mix}$, is related to the
turbulent flame width, $L_f$, or the width of the layer separating the
pure-phase domains. Aiming to study the geometry of the turbulent
flame we measure $L_f$ in the following way: for each point with $1/4
< \theta <3/4$ we find the largest radius of the circle around this
point containing only points with $1/4 <\theta <3/4$; this length,
averaged and multiplied by four, represents the typical width of the
burning region, $L_f$, shown in Fig.~\ref{fig_fspeed}f. We observe
that (i) $L_f$ increases with $\tau$, (ii) $L_f \sim L_\theta$ for
$t<\tau$ (compare Fig.~\ref{fig_fspeed}f and
Fig.~\ref{fig_velocity}c), and (iii) for $t>\tau$, $L_f$ grows with
time but slower than $L_v$, so $L_f/L_v$ decreases with time.

To summarize, we have observed in our numerics the transition at
$t\approx\tau$ from mixed to segregated regimes in variety of ways,
from visual comparison of temperature fields and temperature PDFs to
direct measurements of reaction efficiency and turbulent flame
thickness.

\section{Phenomenology}
\label{Pheno}

This Section focuses on analysis of the asymptotic ($t\gg\tau$),
segregated regime, supposedly characterized by extended inertial
interval and scaling behaviors.  The simulations discussed above do
not run long enough to reach and accurately resolve this
regime. However, the simulations are in a qualitative agreement with
the phenomenology constructed, thus providing a starting point for
more ambitious simulations in the future.

We first review the relevant facts from {\em non-reacting} RT
turbulence, ($\tau=\infty$).  In this case the phenomenology
\cite[]{03Che}, supported by numerical simulations
\cite[]{06CC,05ZWRDB} and some (though limited) experimental
observations, can be summarized as follows.  At $t\gg t_\mathrm{inst}$
the width of the mixing zone grows as $H \propto t^2$.  Velocity
fluctuations within the mixing zone are described by the Kolmogorov
cascade, whereas temperature fluctuations follow the Corrsin-Obukhov
cascade.  Both cascades are driven by the largest RT scale (the spikes
and bubbles) at $L_v\approx L_\theta\propto H$.  Even though this
driving scale grows with time, turbulence at smaller scales is
adiabatically adjusted to the growth. The adiabaticity is due to the
monotonic decrease of the turbulence turnover time with the scale.

As discussed above and illustrated in Fig.~\ref{fig:weak}, for
$t<\tau$ the development in the reacting case proceeds as in the
benchmark non-reacting case. Here $\theta$ is well mixed within the
mixing zone, and the single point distribution of the temperature is
peaked around the median value of $\theta=1/2$. Typical fluctuations
of the velocity and temperature inside the mixing zone (measured in
terms of the differences $\delta_r v$ and $\delta_r\theta$ between two
points separated by a distance $r$) are described by the
Kolmogorov-Corrsin-Obukhov estimates
\begin{eqnarray}
\delta_r v \sim (L_v/t)(r/L_v)^{1/3},
\quad \delta_r\theta \sim (r/L_v)^{1/3},
\quad \mathrm{if}\ L_v\gg r\gg\eta, r_d;
\label{Aeq}
\end{eqnarray}
which are insensitive to the reaction.  The viscous and thermal
diffusion scales, $\eta$ and $r_d$, can be estimated as $\eta\sim
r_d\sim (\nu t)^{3/4}/L_v^{1/2}\propto t^{-1/4}$, provided the Prandtl
number is of order unity.

Since the velocity correlation functions are formed by the direct
cascade determined by the non-linear term in Eq.~(\ref{NSeq}), the
driving effectively occurs at the scale $L_v\approx L_\theta$, with
the former one being not very sensitive to the reaction statistics,
and thus determined by the same estimates as in the non-reacting case,
Eq.~(\ref{Aeq}). (Weak sensitivity of the velocity statistics to the
reaction was also observed in our simulations, see
Fig.~\ref{fig_velocity}.)  In contrast, the $\theta$ statistics are
strongly modified at $t\gg\tau$ when pure phase regions are formed
separated by relatively thin interfaces (see Figs.~\ref{fig_fspeed}
and~\ref{fig_pdfpeaks}).  The interface width $L_f$ is estimated by
balancing advection and reaction terms in Eq.~(\ref{Teq}), $\delta_r
v\delta_r\theta/r\sim \delta_r\theta/\tau$.  One thus derives the
following estimate for the turbulent flame thickness,
\begin{eqnarray} 
   L_f \sim L_\theta (v_{L_\theta}\tau/L_\theta)^{3/2} 
       \sim g \tau^{3/2} t^{1/2}.  
  \label{rG} 
\end{eqnarray} 
The turbulent flame thickness does grow with time, but in such a way
that both $L_\theta/ L_f$ and $L_f/r_d$ also grow as time advances.
(Although our numerical data do not run long enough to illustrate the
increase of $L_f$ with time, we did observe the increase of $L_f$ with
$\tau$.  See Fig.~\ref{fig_fspeed}f.)

Let us now exploit the Kolmogorov-Obukhov relation, calculating the
time derivative of $\theta(\bm r_1)\theta(\bm r_2)$ in accordance with
Eq.~(\ref{Teq}), and then averaging the result over a slice
perpendicular to the $Z$-axis for a fixed value of $\bm r=\bm r_1-\bm
r_2$, where $L_f\ll r\ll L$.  Making use of the adiabaticity and
neglecting the reactive and diffusive terms in
Eqs.~(\ref{NSeq},\ref{Teq}) one derives
\begin{eqnarray}
   \nabla\left\langle(\bm v(\bm r_1)-\bm v(\bm r_2))
   (\theta(\bm r_1)-\langle\theta(\bm r_1)\rangle)
   (\theta(\bm r_2)-\langle\theta(\bm r_2)\rangle)
   \right\rangle \sim \varepsilon_\theta,
   \nonumber
\end{eqnarray}
where the flux term on the right-hand side originates from driving at
the large-scale, $v_z\partial_z\langle\theta\rangle$, due to the large
scale temperature gradient set by buoyancy \cite[]{ShrSig}. Using the
Kolmogorov estimation (\ref{Aeq}) for $\delta_r v$, one arrives at
\begin{equation}
 S_2(r)\equiv\langle(\delta_r\theta)^2 \rangle
 \sim
 \varepsilon_\theta\varepsilon_v^{-1/3} r^{2/3}
 \sim
 (r/H)^{2/3}.
 \label{S2}
\end{equation}
The most important assumption, made while deriving Eq.~(\ref{S2}),
concerns neglecting the reactive contribution into the $\theta^2$
balance. This assumption will be justified later in the Section.

At $r\gg L_f$, Eq.~(\ref{S2}) also describes the probability for two
points separated by $r$ to fall in two distinct domains of
$\theta\approx 0$ and $\theta\approx 1$, or vice versa.  Exactly the
same probabilistic arguments apply to a temperature structure function
of any positive order, thus resulting in the asymptotic independence
of the respective scaling exponent of the order:
\begin{eqnarray}
   S_n(r)\equiv\langle|\delta_r\theta|^n \rangle\sim (r/H)^{2/3}.
   \label{Sn}
\end{eqnarray}
The expression implies strong intermittency of the temperature field,
as the ratios $S_n/(S_2)^{n/2}\sim(H/r)^{(n-2)/3}$ all grow with scale
at $n>2$.  Notice that these arguments also apply to the immiscible
RT, of the type discussed in \cite{05CKL}.

The expressions (\ref{S2},~\ref{Sn}) suggest that the flame interface
is fractal and, moreover, single-fractal (as opposed to multi-fractal,
see e.g. discussion of \cite{89SRM} and \cite{91Ker}).  The
fractalization of the flame is properly explained by dependence of the
fraction, $\Delta_\mathrm{mix}$, of the mixing zone on the turbulence
flame thickness, $L_f$.  Taking Eqs.~(\ref{S2},~\ref{Sn}) at $r\to
L_f$, one estimates, $\Delta_\mathrm{mix}\sim (L_f/H)^{2/3}\sim
\tau/t$.  This also implies that $z_f\sim H$, thus confirming the
general conclusion made in Section~\ref{Prelim}.

Now we are ready to justify the approximation leading to
Eq.~(\ref{S2}).  Let us estimate contribution of the flame/reaction
term, $\mu=\langle[\theta(\bm r_1)-\langle\theta(\bm r_1)\rangle]R(\bm
r_2)\rangle/\tau$, into the aforementioned $\theta^2$ balance
relation.  The product is non-zero only if $\bm r_2$ falls inside the
interface.  Then $\mu$ is much less than $\langle R \rangle/\tau$
since $\theta(\bm r_1)-\langle\theta(\bm r_1)\rangle$ is of order
unity and it also has zero average.  The average $\langle R \rangle$
is determined by the fraction of the well mixed region within the
mixing zone, $\Delta_\mathrm{mix} \sim(L_f/H)^{2/3}$.  One concludes,
that, indeed, $\mu\ll 1/t\sim \varepsilon_\theta$ at $r\gg L_f$, thus
justifying the assumption made above in deriving Eq.~(\ref{S2}) and
subsequently Eq.~(\ref{Sn}).

Turning to discussion of the range of scales smaller than $L_f$ but
larger than $r_d$, one notices that the reaction term does not
contribute to the $\theta^2$ balance, as it is much smaller than the
respective advection contribution.  At these scales, where the direct
effect of the reaction term is irrelevant and the Kolmogorov velocity
estimates still survive, one expects that the Obukhov-Corrsin scaling,
$\delta_r\theta\propto r^{1/3}$, applies. Matching the self-similar
behavior with Eq.~(\ref{Sn}) at $L_f$, one derives the following
scaling relation for the structure functions at $L_f\gg r\gg r_d$
\begin{eqnarray}
   S_n(r)\sim (L_f/H)^{2/3}(r/L_f)^{n/3}.
   \label{S2small}
\end{eqnarray}
This formula allows a simple interpretation: the first term on the
right hand side stands for the fraction of the well mixed region
within the mixing zone, $\Delta_\mathrm{mix}$, where the reaction
takes place, while the second term accounts for the Corrsin-Obukhov
decay of correlations with the decrease in scale. The main
contribution in the RHS of Eq.~(\ref{S2small}) comes from the
interfacial domains, while contributions associated with pure phase
domains, where $\theta\approx 0$ or $\theta\approx 1$, is much
smaller.

\section{Conclusions}

In this manuscript we analyze the reactive RT turbulence in the case
of a slow flame realized when the typical reaction time $\tau$ is
larger then the time of RT instability development, $t_{inst}$.  The
RT instability leads to formation of the mixing zone at $t\sim
t_{inst}$. Further development of the mixing zone is roughly split
into the early turbulent stage, $\tau\gtrsim t\gtrsim t_{inst}$,
characterized by well-mixed hot and cold phases, and the later stage,
$t\gtrsim \tau$, where pure state phases are segregated. In the early
stage, the system is only mildly sensitive to the reaction (flame),
resulting in a slight shift of the mixing zone as a whole, but no
visible feedback of the flame on velocity distribution. The velocity
fluctuations are still insensitive to the reaction, at $t\gtrsim
\tau$, however the temperature fluctuations become modified
significantly. In this regime, burning takes place within thin
interfaces separating large patches of pure phases. Our numerical
simulations confirm the basic prediction of the transition at
$t\approx\tau$. All the qualitative features expected from the
transition are observed in the simulations. Given that our numerical
resources were not sufficient for detailed analysis of the asymptotic,
$t\gg\tau$, stage, we rely here on the phenomenology. Separation of
the advective range for density fluctuations in two ranges, and
formation of a single-fractal flame interface are our two main
phenomenological predictions. This highly intermittent behavior of
density fluctuations is in a great contrast with what was observed
without reaction. These and other predictions of the phenomenology are
subjects for juxtaposition with future numerical and laboratory
experiments.

Notice that the general picture of mixing at $t>\tau$ discussed above
is reminiscent of the immiscible turbulence discussed in
\cite{05CKL}. In both cases large regions of pure phases are
separated by thin interfaces.  Viewed at the scales larger than the
interface width, the interfaces are advected passively.  Moreover,
strong intermittency in the density fluctuations, reported above for
the reactive case, also takes place in the setting of externally
stirred (e.g. by gravity) immiscible turbulence.

Finally, our phenomenology also applies to other regimes of turbulent
flames, e.g. those realized in combustion engines, where the turbulent
flame width is positioned in between the integral scales of turbulence
and the viscous/diffusive scales, $L_v,L_\theta\gg L_f\gg\eta,r_d$.

\section{Acknowledgments}
We wish to thank P.~Fischer for the permission to use the Nekton code,
A.~Obabko and P.~Fischer for the detailed help in using the code, and
V.G.~Weirs for useful comments. This work was supported by the
U.S. Department of Energy at Los Alamos National Laboratory under
Contract No. DE-AC52-06NA25396, under Grant No.\ B341495 to the Center
for Astrophysical Thermonuclear Flashes at the University of Chicago,
and RFBR grant 06-02-17408-a at the Landau Institute.

\bibliographystyle{jfm.bst}

\bibliography{rrtt}

\end{document}